# Effect of the interface resistance in non-local Hanle measurements

Estitxu Villamor,[1] Luis E. Hueso,[1,2] and Fèlix Casanova,[1,2,*]

[1]CIC nanoGUNE, 20018 Donostia-San Sebastian, Basque Country (Spain)
[2]IKERBASQUE, Basque Foundation for Science, 48011 Bilbao, Basque Country (Spain)

*E-mail: f.casanova@nanogune.eu

We use lateral spin valves with varying interface resistance to measure non-local Hanle effect in order to extract the spin-diffusion length of the non-magnetic channel. A general expression that describes spin injection and transport, taking into account the influence of the interface resistance, is used to fit our results. Whereas the fitted spin-diffusion length value is in agreement with the one obtained from standard non-local measurements in the case of a finite interface resistance, in the case of transparent contacts a clear disagreement is observed. The use of a corrected expression, recently proposed to account for the anisotropy of the spin absorption at the ferromagnetic electrodes, still yields a deviation of the fitted spin-diffusion length which increases for shorter channel distances. This deviation shows how sensitive the non-local Hanle fittings are, evidencing the complexity of obtaining spin transport information from such type of measurements.

## I.  INTRODUCTION

Pure spin currents are a key ingredient in the field of *spintronics* [1], which takes advantage not only of the charge of the electron, but also of its spin as an alternative to transport information. Lateral spin valves (LSVs), consisting of two ferromagnetic (FM) electrodes bridged by a non-magnetic (NM) channel (see Fig. 1(a)), are widely used to electrically create pure spin currents due to their non-local geometry, in which a spin-polarized current is injected from one of the FM electrodes (the injector) into the NM channel, and the pure spin current at the second FM electrode (the detector) is measured [2-15].

Hanle effect is based on the precession of spins under a perpendicular magnetic field. Due to the diffusive nature of the spin transport through the NM, there is dispersion on the time that spins need to travel from the FM injector to the detector, which in turn originates an angular dispersion on the orientation of the spins arriving at the FM detector. This causes the measured spin current at the FM detector to be zero for high enough magnetic fields [3-9]. In addition to being an effective tool for spin manipulation, it presents an important advantage in the study of the spin-injection and transport mechanisms, because it permits to obtain the spin polarization of the FM ($P_F$), of the FM/NM interface ($P_I$) and the spin-diffusion length of the NM ($\lambda_N$) by using a single LSV [3-8], as opposed to the conventional non-local spin valve (NLSV) method, which needs several LSVs with different distances ($L$) between the FM electrodes in order to obtain these parameters [10-15]. However, Hanle measurements are very sensitive to different device details, such as the interface resistance [7,8] or the finite length of the NM channel [9]. The used model has also been widely discussed in terms of the liability of the obtained information. It has been suggested that it is not possible to measure Hanle effect with transparent interfaces [3,12] or that, if doing so, the equation needs to be carefully chosen [7,8].

In the present work, we analyze the validity of the general expression for the study of spin injection and transport in LSVs with any FM/NM interface resistance, presented from Ref. 5. We do so by fitting the equation to measurements of the Hanle effect in LSVs with different interface resistances and comparing the obtained parameters to those obtained from the fitting of the NLSV measurements as a function of $L$ in the very same devices. Whereas in the presence of a contact resistance both methods are in good agreement, we observe an anomalous behaviour for the case with transparent contacts, where there is a clear mismatch between both methods. While, for $L$ larger than $\lambda_N$, this disagreement can be solved by taking into account the recently proposed spin absorption anisotropy at the FM electrodes [8], it is still present when $L$ is shorter than $\lambda_N$, evidencing that an additional effect is influencing the spin precession. Our analysis shows the complexity of an accurate fitting of non-local Hanle measurements, a widely used technique to extract relevant spin-transport parameters.

## II. EXPERIMENTAL DETAILS

The LSVs employed in this work were fabricated by a two-step electron-beam lithography, ultra-high-vacuum (base pressure $\leq 1\times10^{-8}$mbar) evaporation and lift-off process. In the first step, FM electrodes were patterned in PMMA resist on top of a Si/SiO$_2$ substrate and 35nm of permalloy (Py) or cobalt (Co) were evaporated. Different widths of the FM electrodes were chosen, $w_{F1}$~85nm and $w_{F2}$~140nm, in order to obtain different switching magnetic fields. In the second step, the NM channel with a width of $w_N$~190nm was patterned and Cu was thermally evaporated with a thickness $t$ ~150nm. Ar-ion milling was performed prior to the Cu deposition in order to remove resist leftovers [14]. The reason for choosing different materials as FM electrodes is the need of different FM/NM interface resistances. Py has given us high-quality transparent interfaces with a high spin polarization [13,14], whereas Co is easily oxidized allowing the fabrication of an interface with a non-zero resistance [15]. The interface resistance ($R_I$) was measured in all samples, where a cross-shaped junction was fabricated in addition to the regular LSVs. Several samples were fabricated and measured (all of them containing LSVs with different $L$). Since the obtained results are reproducible [16], only two samples will be compared in this paper. Sample #1, containing Co/Cu LSVs, has an $R_I \times A_I$ product ($A_I$ is the contact area) of $2.8\times10^{-2}\Omega\mu m^2$ (the $R_I$'s have values of $R_{I1}$=1.6$\Omega$ and $R_{I2}$=1$\Omega$, which fall in the intermediate regime, *i.e.* they are not transparent interfaces but they cannot be considered to be in the fully tunneling regime [17]). The measured $R_I$ at the Py/Cu junctions of sample #2 is negative, meaning that $R_I$ is of the order or lower than the resistance of the electrodes and $R_I \times A_I \leq 10^{-3}\Omega\mu m^2$ [14,18,19]. Therefore, sample #2 is in the transparent regime [14,17].

All measurements were performed in a liquid He cryostat at 10 K, applying a magnetic field $B$ and using a "DC-reversal" technique [11]. The voltage $V$, normalized to the applied current $I$, is defined as the non-local resistance $R_{NL}$=$V/I$ (see Fig. 1(a) for a scheme of the measurement). This magnitude is positive [negative] when the magnetization of the electrodes is parallel (P) [antiparallel (AP)], depending on the value of $B$. Two types of measurements have been performed: (i) $R_{NL}$ as a function of the in-plane magnetic field along the FM electrodes ($B_Y$ from Fig. 1(a)), so-called *NLSV* measurements, and (ii) $R_{NL}$ as a function of the out-of-plane magnetic field ($B_Z$ from Fig. 1(a)), so-called *Hanle* measurements. In the case of NLSV measurements, the absolute value of $R_{NL}$ does not vary, only its sign does change when the magnetizations of the FM electrodes change from P to AP. The difference between the positive and the negative values of $R_{NL}$ is the spin signal, $\Delta R_{NL}$=2×$R_{NL}$, which is proportional to the spin accumulation at the FM detector (see lower inset of Fig. 1(b)). In the case of Hanle measurements, the magnitude of the measured $R_{NL}$ gradually changes from positive to negative (or *vice versa*) due to the precession of the spins. In addition, a reduction in $R_{NL}$ with $B_Z$ is superimposed, due to the angular dispersion of the orientation of the spins [6].

The expression used for fitting the Hanle measurements, obtained by solving the Bloch-type equation with an added one-dimensional spin-diffusion term applied to the LSV geometry [2,5,10,20], is the following:

$$R_{NL} = \frac{2\tilde{R}_N \left[\frac{P_I}{1-P_{I1}^2}\frac{R_{I1}}{\tilde{R}_N} + \frac{P_F}{1-P_F^2}\frac{R_{F1}}{\tilde{R}_N}\right]\left[\frac{P_I}{1-P_{I2}^2}\frac{R_{I2}}{\tilde{R}_N} + \frac{P_F}{1-P_F^2}\frac{R_{F2}}{\tilde{R}_N}\right]\left(\frac{Re[\tilde{\lambda}_N e^{-L/\tilde{\lambda}_N}]}{Re[\tilde{\lambda}_N]}\right)}{\left[1+\frac{2}{1-P_{I1}^2}\frac{R_{I1}}{\tilde{R}_N} + \frac{2}{1-P_F^2}\frac{R_{F1}}{\tilde{R}_N}\right]\left[1+\frac{2}{1-P_{I2}^2}\frac{R_{I2}}{\tilde{R}_N} + \frac{2}{1-P_F^2}\frac{R_{F2}}{\tilde{R}_N}\right] - \left(\frac{Re[\tilde{\lambda}_N e^{-L/\tilde{\lambda}_N}]}{Re[\tilde{\lambda}_N]}\right)^2}, \quad (1)$$

where $\tilde{\lambda}_N = \frac{\lambda_N}{\sqrt{1+i\omega_L\tau_{sf}}}$ and $\tilde{R}_N = Re[\tilde{\lambda}_N]\rho_N/t_N w_N$ are an effective spin-diffusion length and an effective spin resistance of the NM, respectively, and $R_{Fi}=\lambda_F\rho_F/w_N w_{Fi}$ is the spin resistance of the FM injector ($i=1$) or detector ($i=2$). $\lambda_F$ is the spin-diffusion length of the FM, $\rho_N$ and $\rho_F$ are the electrical resistivities of the NM and FM, $\tau_{sf}$ is the spin-relaxation time of the NM and $\omega_L=2\mu_B B_Z/\hbar$ is the Larmor frequency, with $\mu_B$ the Bohr magneton and $\hbar$ the reduced Planck constant. $\rho_{Cu}(=1.2\mu\Omega cm)$ is obtained by measuring the resistance of Cu for every $L$, and performing a linear fit for each sample, whereas $\rho_{Py}(=22.4\mu\Omega cm)$ and $\rho_{Co}(=11.5\mu\Omega cm)$ are obtained in two different devices, where Py and Co were deposited under the same nominal conditions as for the LSVs. We use $\lambda_{Py}=5$nm [21] and $\lambda_{Co}=36$nm [21]. The dimensions $w_N$, $w_{Fi}$ and $L$ are measured by Scanning Electron Microscopy (SEM) for each device. Therefore, $P_I$, $P_F$ and $\lambda_{Cu}$ are the parameters to be fitted from Hanle measurements. To be more precise, one needs to take into account that the magnetization of the FM electrodes can be tilted out-of-plane due to $B_Z$. The following equation is used to correct for such tilting [3,4,6]:

$$R_{NL}^{P(AP)}(B_Z,\theta) = \pm R_{NL}^P(B_Z)\cos^2\theta + |R_{NL}(B_Z=0)|\sin^2\theta, \quad (2)$$

where "+" and "-" signs correspond to the P and AP magnetizations of the FM electrodes, $R_{NL}^P(B_Z)$ is the one from Eq. (1), and $\theta\equiv\theta(B_Z)$ is the angle between the magnetization of the FM electrodes and $B_Z$; its dependence with $B_Z$ can be extracted from the anisotropic magnetoresistance (AMR) measurements of the FM electrodes as a function of $B_Z$ [6]. Hence, in order to obtain the spin polarizations and spin-diffusion length from the Hanle measurements, the data was fitted to Eq. (2) (see upper inset of Fig. 1(b)).

In the case of NLSV measurements we have an in-plane magnetic field $B_Y$, and Eq. (1) reduces to the following:

$$R_{NL} = \frac{2R_N \left[\frac{P_I}{1-P_{I1}^2}\frac{R_{I1}}{R_N} + \frac{P_F}{1-P_F^2}\frac{R_{F1}}{R_N}\right]\left[\frac{P_I}{1-P_{I2}^2}\frac{R_{I2}}{R_N} + \frac{P_F}{1-P_F^2}\frac{R_{F2}}{R_N}\right]e^{-L/\lambda_N}}{\left[1+\frac{2}{1-P_{I1}^2}\frac{R_{I1}}{R_N} + \frac{2}{1-P_F^2}\frac{R_{F1}}{R_N}\right]\left[1+\frac{2}{1-P_{I2}^2}\frac{R_{I2}}{R_N} + \frac{2}{1-P_F^2}\frac{R_{F2}}{R_N}\right] - e^{-2L/\lambda_N}}, \quad (3)$$

where $R_N=\lambda_N\rho_N/t_N w_N$ and $\lambda_N$ are the regular spin resistance and spin-diffusion length of the NM metal, respectively. The measured $\Delta R_{NL}$ as a function of $L$ can, thus, be fitted to Eq. (3) (see Fig. 1 (b)). Even though the values obtained from both methods should be identical, the validity of Hanle measurements in the case of transparent contacts has already been called into question [3,7,8,12].

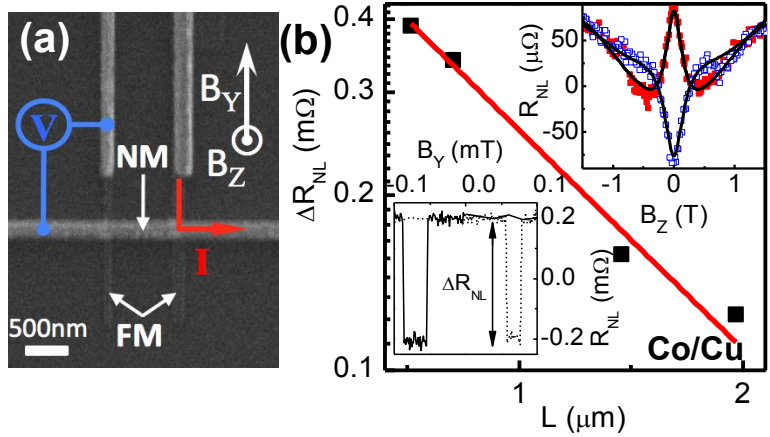

Figure 1: (a) SEM image of a LSV. The non-local measurement configuration, materials, and the directions of the in-plane and out-of-plane magnetic fields ($B_Y$ and $B_Z$) are shown. (b) Spin signal, $\Delta R_{NL}$, as a function of the distance between FM electrodes, $L$, measured at 10 K for sample #1, which contains Co/Cu LSVs with an interface resistance of ~1Ω. Red solid line is a fit to Eq. (3). Lower inset: non-local resistance, $R_{NL}$, as a function of $B_Y$ measured at 10 K for the same Co/Cu LSV with $L$=500 nm. Solid (dotted) line indicates the decreasing (increasing) sweep of $B_Y$. $\Delta R_{NL}$ is tagged in the image. Upper inset: $R_{NL}$ as a function $B_Z$ measured at 10 K both for the parallel (red solid squares) and anti-parallel (blue empty squares) configuration of the FM electrodes for a Co/Cu LSV with $L$=1.5μm. Black solid lines are fits to Eq. (2), using the $R_{NL}$ expression from Eq. (1).

## III. RESULTS AND DISCUSSION

For sample #1, with a non-zero interface resistance, $P_I^{NLSV}$=0.043±0.003, $P_{Co}^{NLSV}$=0.038±0.004 and $\lambda_{Cu}^{NLSV}$=1159±100nm were obtained from the fitting of the NLSV measurements to Eq. (3). The measured data and the fitting are shown in Fig. 1(b). The value of $\lambda_{Cu}^{NLSV}$ is in good agreement with our previous results [13,14], whereas the low value of $P_{Co}^{NLSV}$ has also been reported and discussed before [10,14]. Note that $P_I$ and $P_F$ are coupled, as seen from Eqs. (1)-(3), since sample #1 is not fully in the tunnelling regime. Only when $\frac{P_I R_{Ii}}{1-P_I^2} \gg \frac{P_F R_{Fi}}{1-P_F^2}$ or $\frac{P_I R_{Ii}}{1-P_I^2} \ll \frac{P_F R_{Fi}}{1-P_F^2}$ (*i.e.* for the tunnelling or transparent regimes [17]) they will decouple.

For Hanle measurements, $R_{NL}$ as a function of $B_Z$ was measured for both P and AP magnetization states (see inset of Fig. 1(b)), with identical results. For all the LSVs with different $L$, a spin-diffusion length ranging between $\lambda_{Cu}^{Hanle}$=987±25nm and 1107±27nm, and an interface polarization ranging between $P_I^{Hanle}$=0.044±0.001 and 0.048±0.001 were obtained. Due to the coupling of $P_I$ and $P_F$ in Eq. (1), the spin polarization of Co was fixed to $P_{Co}$=0.038. The obtained $\lambda_{Cu}^{Hanle}$ and $P_I^{Hanle}$ values show no substantial deviation from the NLSV values for any of the distances $L$ (see Fig. 2a).

For sample #2, with transparent interfaces, we can approximate $R_I$=0 in Eqs. (1)-(3) in order to obtain $P_{Py}$ and $\lambda_{Cu}$. From NLSV measurements as a function of $L$ we obtained $P_{Py}^{NLSV}$=0.36±0.01 and $\lambda_{Cu}^{NLSV}$=1125±62nm. However, for Hanle measurements, spin-diffusion lengths ranging between $\lambda_{Cu}^{Hanle}$=557±26nm and 1245±58nm were obtained. The spin polarization of Py also changed between $P_{Py}^{Hanle}$=0.34±0.01 and 0.63±0.02. Note that in this case $R_{NL}$ as a function of $B_Z$ was only measured for the P magnetization of the FM electrodes [22]. As shown in Fig. 2a, the obtained $\lambda_{Cu}^{Hanle}$ values present a clear deviation from the NLSV values with a strong dependence on $L$: for low values of $L$ ($L \ll \lambda_{Cu}^{NLSV}$) the agreement between both methods is excellent but, as $L$ increases, $\lambda_{Cu}^{Hanle}$ starts to deviate from $\lambda_{Cu}^{NLSV}$. The highest discrepancy occurs for $L \sim \lambda_{Cu}^{NLSV}$ and, for longer $L$ ($L \gg \lambda_{Cu}^{NLSV}$), the deviation of $\lambda_{Cu}^{Hanle}$ tends to reduce. $P_{Py}^{Hanle}$ changes with the opposite tendency to that of $\lambda_{Cu}^{Hanle}$, showing a coupling between both fitting parameters (Fig. 2b). The observed deviation for $L \sim \lambda_{Cu}^{NLSV}$ is clearly originated from a bad fitting of the data [16]. However, this deviation is very reproducible for all measured samples and, thus, intrinsic to LSVs with transparent contacts [16]. Figure 3 shows the measured $R_{NL}$ as a function of $B_Z$ in sample #2 for the three mentioned regimes, together with the simulated curves of Eq. (2), using the $R_{NL}$ expression from Eq. (1). For the simulations (blue solid lines), we used the $P_{Py}^{NLSV}$ and $\lambda_{Cu}^{NLSV}$ values obtained from the fittings of the NLSV measurements. The figure shows a good agreement between the measured data and Eq. (1) for $L \ll \lambda_{Cu}^{NLSV}$, the same way there is an excellent agreement between the fitted $\lambda_{Cu}^{Hanle}$ and $\lambda_{Cu}^{NLSV}$. However, in the $L \sim \lambda_{Cu}^{NLSV}$ regime, the curves are far from reproducing the measured data. For the $L \gg \lambda_{Cu}^{NLSV}$ regime, the simulated curve tend to converge to the measured data again. This result suggests that Eq. (2) (with the $R_{NL}$ from Eq. (1)) is not valid and additional effects should be considered in the spin transport in Cu.

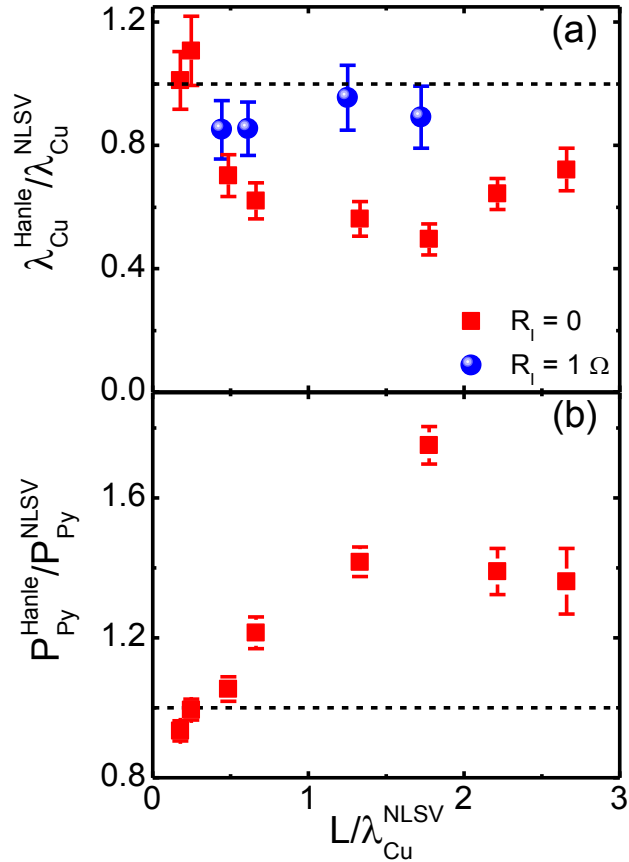

Figure 2: (a) Spin-diffusion length of Cu ($\lambda_{Cu}^{Hanle}$) obtained from the fitting of Eq. (2) (using $R_{NL}$ from Eq. (1)) to the $R_{NL}$ vs. $B_Z$ data, as a function of $L$, for sample #2 containing Py/Cu LSVs with transparent interfaces (red solid squares) and sample #1 containing Co/Cu LSVs with an interface resistance of ~1Ω (blue solid circles). Both $\lambda_{Cu}^{Hanle}$ and $L$ are normalized to the spin-diffusion length of Cu ($\lambda_{Cu}^{NLSV}$) obtained for each sample from the fitting of Eq. (3) to the $\Delta R_{NL}$ vs. $L$ data. (b) Spin polarization of Py ($P_{Py}^{Hanle}$) obtained from the same fitting of Eq. (2) (using $R_{NL}$ from Eq. (1)) to the $R_{NL}$ vs. $B_Z$ data, as a function of $L$, for sample #2. $P_{Py}^{Hanle}$ is normalized to the spin polarization of Py ($P_{Py}^{NLSV}$) obtained for the same sample from the fitting of Eq. (3) to the $\Delta R_{NL}$ vs. $L$ data.

Whereas Maasen *et al.* reported an anomalous behaviour of the parameters obtained from Hanle measurements due to a bad fitting, where the backflow of spins at the FM electrodes was not taken into account [7], this is not the case in the present work, since Eq. (1) explicitly takes into account the role of the interface resistances. Very recently, Idzuchi and co-workers [8] have proposed the difference in the spin absorption mechanisms for longitudinal and transverse spin currents as the reason of the disagreement in Hanle measurements in LSVs without tunnel barriers. According to this work, in LSVs with transparent interfaces, the different spin absorption by the FM electrodes for different current polarizations alters the spatial distribution of the chemical potential. Therefore, the spin transport is also altered, more pronouncedly for short $L$ [8]. This could explain the strong deviation between $\lambda_{Cu}^{Hanle}$ and $\lambda_{Cu}^{NLSV}$ in the $L \sim \lambda_{Cu}^{NLSV}$ regime, but one would expect an even stronger deviation in the $L << \lambda_{Cu}^{NLSV}$ regime. Instead, we find the opposite trend.

In order to clarify this issue, Fig. 3 also shows the simulated curves of Eq. (2), using now the $R_{NL}$ expression from Eq. (S13) in Ref. 8 (red dashed lines). For the simulations, in addition to the $P_{Py}^{NLSV}$ and $\lambda_{Cu}^{NLSV}$ values obtained from the fittings of the NLSV measurements, a value of $G_r=3.9\times10^{14}\Omega^{-1}m^{-2}$ was used as the real part of the spin-mixing conductance of the Py/Cu interface [8,23,24]. For the $L>>\lambda_{Cu}^{NLSV}$ regime, Eq. (S13) from Ref. 8 follows quite accurately the measured data. However, in the $L \sim \lambda_{Cu}^{NLSV}$ regime, the simulated curves start to deviate from the experimental results. The discrepancy is highest for the $L<<\lambda_{Cu}^{NLSV}$ regime, where the measured data is more affected by the precession, suggesting that the diffusion time is longer, an effect already reported to alter the fitted $P_F$ in LSVs using Eq. (3) [11].

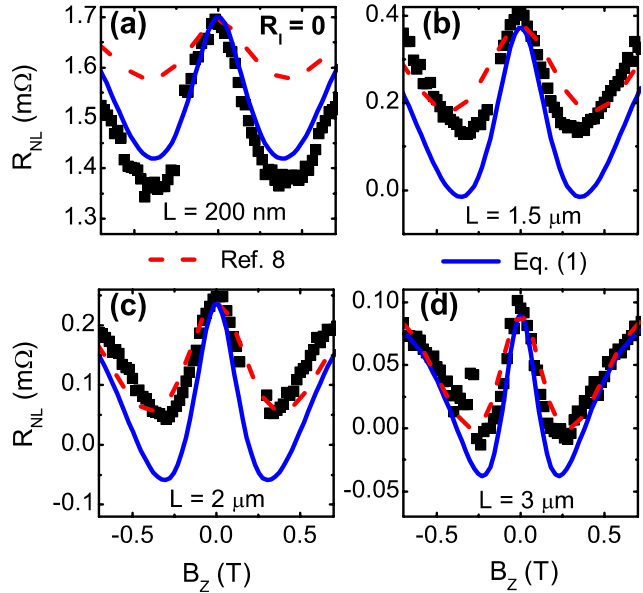

Figure 3: $R_{NL}$ measured as a function of $B_Y$ (black squares) for sample #2. $L$ ranges from 200nm to 3μm. All measurements were done for a parallel configuration of the Py electrodes at 10 K. Blue solid (red dashed) line is a simulation of Eq. (2) using $R_{NL}$ from Eq. (1) (Eq. (S13) from Ref. 8). $P_{Py}^{NLSV}$ and $\lambda_{Cu}^{NLSV}$ obtained from NLSV measurements were used, and a real part of the spin-mixing conductance between Py and Cu of $G_r=3.9\times10^{14}\Omega^{-1}m^{-2}$ was assumed [8,23,24].

In order to obtain the value of $\lambda_{Cu}$ by fitting Eq. (2) with $R_{NL}$ from Ref. 8, we fixed all the parameters except for $\lambda_{Cu}^{Hanle}$, which was left as the fitting parameter. This was done for the sake of simplicity, given the complexity of Eq. (S13) from Ref. 8.

Figure 4 shows the obtained values of $\lambda_{Cu}^{Hanle}$ as a function of $L$ using that equation. For comparison, the $\lambda_{Cu}^{Hanle}$ values obtained by using Eq. (1), already shown in Fig. 2a, are also plotted. The tendency is the same observed in the simulations, where $\lambda_{Cu}^{Hanle}$ and $\lambda_{Cu}^{NLSV}$ are in good agreement in the $L>>\lambda_{Cu}^{NLSV}$ regime, but $\lambda_{Cu}^{Hanle}$ decreases when $L<<\lambda_{Cu}^{NLSV}$. Therefore, Eq. (S13) from Ref. 8, which considers both the spin backflow and the anisotropic spin absorption at the FM/NM interfaces, does not work at the $L<<\lambda_{Cu}^{NLSV}$ regime, showing that both mentioned effects are not enough to account for the disagreement between the current Hanle models and the measured curves.

A possible source of interference is the effect of nearby FM electrodes in the LSVs, but it is discarded by performing control experiments [16,25]. Taking into account that the discrepancy occurs at short channel distances (see green triangles in Fig. 4), the origin could be attributed to the use of a one-dimensional spin-diffusion model to derive the used equations [5,8], which could no longer be a good approximation. Indeed, the region of the NM channel under the FM injector, where the spin-polarized electrons spend time diffusing, has been shown to influence the effective spin polarization of the FM in LSVs [11] and would also affect the non-local Hanle curves [26].

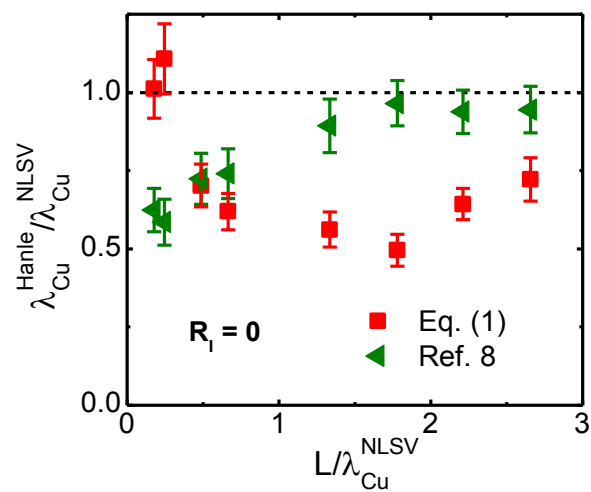

Figure 4: $\lambda_{Cu}^{Hanle}$ obtained from the fitting of Eq. (2) by using Eq. (1) (red squares) and Eq. (S13) from Ref. 8 (green tringles) as a function of $L$ for sample #2, which consists of Py/Cu LSVs with transparent interfaces. Both $\lambda_{Cu}^{Hanle}$ and $L$ are normalized to $\lambda_{Cu}^{NLSV}$.

## IV. CONCLUSIONS

To summarize, we performed non-local Hanle measurements in LSVs with transparent and finite interface resistances, and we compared the spin-diffusion length of Cu, $\lambda_{Cu}$, obtained from such measurements to the one obtained from NLSV measurements as a function of $L$. Whereas, in the case where we have a finite FM/NM interface resistance, both methods are in excellent agreement, in the case of transparent interfaces an anomalous behaviour is observed, which depends on the distance $L$ between both FM electrodes. Although taking into account the spin backflow and the anisotropic spin absorption at the FM/NM interfaces can explain some of the observed disagreements, an additional interference that influences the non-local Hanle measurements is detected when $L<<\lambda_{Cu}$. Such effect is beyond the understanding of the current one-dimensional spin diffusion models, evidencing the need for a more complete model that takes into account three dimensional effects. Hence, care should be taken when obtaining spin-transport information from such type of measurements in LSVs with transparent interfaces.

**ACKNOWLEDGEMENTS**

The authors thank Asier Ozaeta and F. Sebastián Bergeret for fruitful discussions. This work was supported by the European Union 7th Framework Programme under the Marie Curie Actions (256470-ITAMOSCINOM) and the European Research Council (257654-SPINTROS), and by the Spanish MINECO under Project No. MAT2012-37638. E. V. thanks the Basque Government for a PhD fellowship (Grant No. BFI-2010-163).

**Effect of the interface resistance in non-local Hanle measurements**
Estitxu Villamor,[1] Luis E. Hueso,[1,2] and Fèlix Casanova[1,2]

[1]CIC nanoGUNE, 20018 Donostia-San Sebastian, Basque Country (Spain)
[2]IKERBASQUE, Basque Foundation for Science, 48011 Bilbao, Basque Country (Spain)

**SUPPLEMENTAL MATERIAL**

### S1. Hanle fittings from sample #2

Figure S1 shows the measured $R_{NL}$ as a function of $B_Z$ in sample #2, as well as the fittings to Eq. (2) (using $R_{NL}$ from Eq. (1)), for three different regimes: (*i*) $L \ll \lambda_{Cu}^{NLSV}$ (Fig. S1(a)), (*ii*) $L \sim \lambda_{Cu}^{NLSV}$ (Figs. S1(b) and S1(c)), (*iii*) $L \gg \lambda_{Cu}^{NLSV}$ (Fig. S1(d)). The agreement between the measured data and the fitted curve is only good for the first regime, where the values of $\lambda_{Cu}^{Hanle}$ and $\lambda_{Cu}^{NLSV}$ are also in good agreement. For the $L \gg \lambda_{Cu}^{NLSV}$ regime, $\lambda_{Cu}^{Hanle}$ and $\lambda_{Cu}^{NLSV}$ tend to be similar again. As seen from this figure, in the intermediate regime the fitted curve tends to be wider than the measured data, which decreases considerably the fitted value of $\lambda_{Cu}^{Hanle}$ (increasing, in turn, the value of $P_{Py}^{Hanle}$).

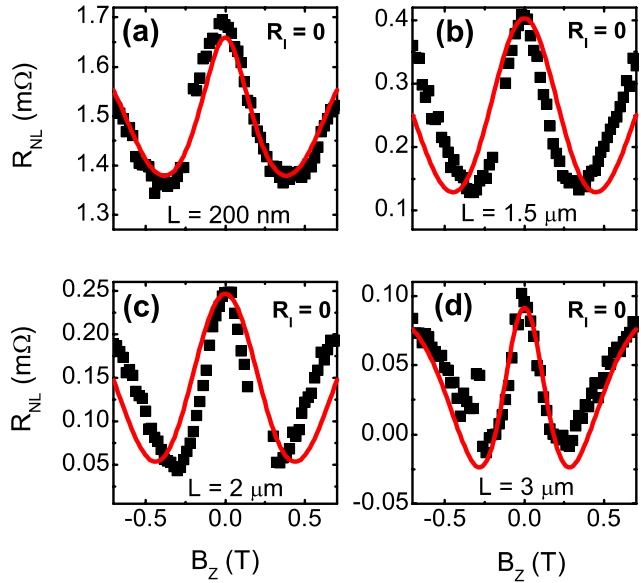

Figure S1: Non-local resistance $R_{NL}$ measured as a function of $B_Z$ (black squares) for sample #2, which consists of Py/Cu LSVs with transparent interfaces. $L$ ranges from 200 nm to 3 μm. Red solid lines are fits to Eq. (2) (using $R_{NL}$ from Eq. (1)).

### S2. Additional samples

In the main text, only two samples are compared for the sake of clarity: sample #1 (containing Co/Cu LSVs with a non-zero interface resistance) and sample #2 (containing Py/Cu LSVs with transparent interfaces). However, as mentioned in the text, more samples were measured. The results from these extra samples are shown in this section in order to emphasize that samples with the same interface resistance have the same behavior. Figure S2 shows the spin diffusion length of copper (Cu) obtained from

Hanle measurements, $\lambda_{Cu}^{Hanle}$, as a function of the distance $L$ between FM electrodes; both quantities are normalized to the spin diffusion length of Cu obtained from NLSV measurements, $\lambda_{Cu}^{NLSV}$, for six different samples. For details about the measurements and how the spin diffusion length is obtained we refer the reader to the main text. It is observed that in samples with a non-zero interface resistance (Fig. S2(a)) both methods are in good agreement (*i.e.*, $\lambda_{Cu}^{Hanle}$ does not deviate significantly from $\lambda_{Cu}^{NLSV}$), whereas in samples with transparent interfaces (Fig. S2(b)) there is a strong deviation which depends on $L$, and has the same trend for all the measured samples.

In addition, the effect of the nearby electrodes is considered as a possible source of interference, due to the design of our devices, which consist of several LSVs on a row. However, by systematically varying the distance of the nearby Py/Cu LSVs with transparent interfaces, the same behavior is observed, ruling out any effect coming from the adjacent electrodes. Two of the control samples where the distance between the Py electrodes was varied are shown in Fig. S2(b).

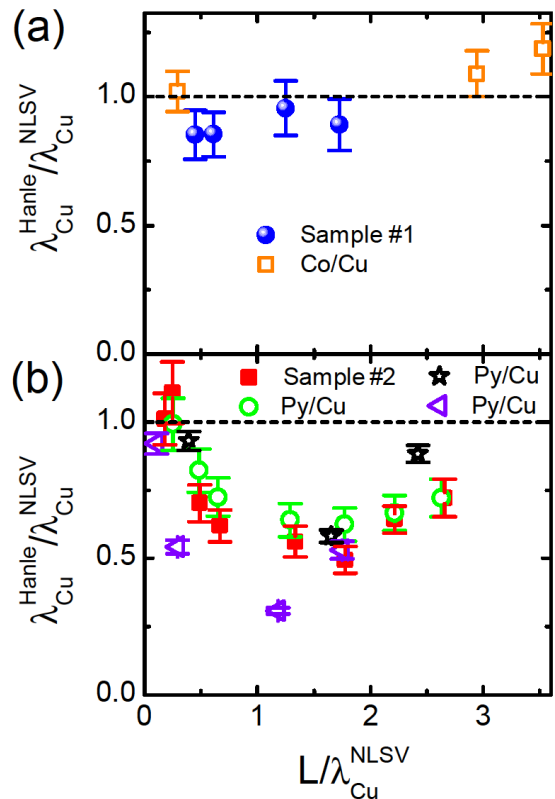

Figure S2: $\lambda_{Cu}^{Hanle}$ as a function of $L$ for (a) Co/Cu LSVs with $R_I \sim 1\Omega$ and (b) Py/Cu LSVs with transparent interfaces. Both $\lambda_{Cu}^{Hanle}$ and $L$ are normalized to the $\lambda_{Cu}^{NLSV}$ value of each sample.